\documentclass{agujournal2018}
\usepackage{apacite}
\usepackage{url} 

\usepackage{graphicx}
\usepackage{amsmath}
\usepackage[varg]{txfonts}
\usepackage{xspace}
\usepackage{siunitx}
\usepackage[version=3]{mhchem}
\usepackage[utf8]{inputenc}

\newcommand{\NtwoB}{\ensuremath{\mathrm{N}_2(B^3\Pi_g)}\xspace}
\newcommand{\NtwoC}{\ensuremath{\mathrm{N}_2(C^3\Pi_u)}\xspace}

\definecolor{AL}{rgb}{1.0,0,0}

\draftfalse

\journalname{Geophysical Research Letters}

\begin{document}

\title{On the emergence mechanism of carrot sprites}

\authors{A. Malagón-Romero\affil{1}, J. Teunissen\affil{2},
H.C. Stenbaek-Nielsen\affil{4}, M.G. McHarg\affil{5}, U. Ebert\affil{2,3}, A. Luque\affil{1}}

\affiliation{1}{IAA-CSIC, Granada, Spain}
\affiliation{2}{Centrum Wiskunde \& Informatica (CWI), Amsterdam, The Netherlands}
\affiliation{3}{Eindhoven University of Technology, Eindhoven, The Netherlands}
\affiliation{4}{Geophysical Institute, University of Alaska Fairbanks, 903 Koyukuk Drive, Fairbanks,
Ak 99775-7320, USA}
\affiliation{5}{Department of Physics, 2354 Fairchild Drive, Suite 2A31, US Air Force Academy, CO 80840, USA}

\correspondingauthor{A. Malagón-Romero}{amaro@iaa.es}

\begin{keypoints}
\item Sprite glows emerge as bright and sharply defined regions with low conductivity and an enhanced electric field.
\item Upward negative streamers in carrot sprites emerge predominantly close to the lower boundary of the glow.
\item The upward streamer channel develops a secondary glow with its upper boundary positively charged.
\end{keypoints}

\begin{abstract}
We investigate the launch of negative upward streamers from sprite glows.
This phenomenon is readily observed in high-speed observations of sprites and underlies the classification of sprites into carrot or column types. First, we describe how an attachment instability leads to a sharply defined region in the upper part of the streamer channel. This region has an enhanced electric field, low conductivity and strongly emits in the first positive system of molecular nitrogen. We identify it as the sprite glow. We then show how, in the most common configuration of a carrot sprite, several upward streamers emerge close to the lower boundary of the glow, where negative charge gets trapped and the lateral electric field is high enough. These streamers cut off the current flowing towards the glow and lead to the optical deactivation of the glow above. Finally, we discuss how our results naturally explain angel sprites.

\end{abstract}

\section*{Plain Language Summary}
\noindent
Sprites are large electric discharges that develop in the mesosphere. Sprites often start as one channel that later branches, leading to an intricate structure of hundreds of filaments. According to their morphology, most sprites are classified into column and carrot types. The most noticeable difference is that electrical discharge fronts in column sprites
propagate exclusively downwards whereas carrot sprites also shoot upward propagating channels. Observations reveal that these upward channels emerge from glowing structures in the main channel. In this work we study the evolution of a sprite channel and propose a mechanism for the emergence of carrot sprites and angel sprites, a rare sprite variety that presents an inverted v-shape accompanying the main downward channel.

\section{Introduction}
First reported by \citet{Franz1990/Sci}, sprites are filamentary electric discharges that develop tens of kilometers above thunderclouds (55-80 km altitude). Early work pointed out that sprite filaments are streamers \citep{Pasko1996/GeoRL,Pasko1998/GeoRL,Raizer1998/JPhD}, as was later confirmed by high-speed observations \citep{Stanley1999/GeoRL,Cummer2006/GeoRL,McHarg2007/GeoRL,Stenbaek-Nielsen2008/JPhD}. Nowadays it is accepted that sprites originate from a quasi-electrostatic field produced by uncompensated electric charges due to lightning in the troposphere \citep{Pasko1995/GeoRL,Pasko1997/JGR}. Several works \citep{Luque2009/NatGe,Liu2015/NatCo/2,kohn2019/JGRSP,wu2019/PlasmaRes} point out to the presence of inhomogeneities as a requirement for the initiation sprite streamers. 

After the streamer head passage, the streamer wake develops intricate luminous patterns
in the form of columns and spots known as glows and beads respectively \citep{Stenbaek-Nielsen2008/JPhD}. Here we adopt the term "glow" following previous works by \citet{Stenbaek-Nielsen2013/SGeo,Luque2016/JGRA}. Note that this definition of glow is different from that used by e.g. \citet{Bor2013/JASTP}, where it denotes a diffuse brightness surrounding the upper regions of a sprite.
In our context, glows fade out on time scales spanning from a few milliseconds to a few hundred milliseconds and often with similar decay times \citep{Luque2016/JGRA}, which indicates a tight coupling
between different regions of a sprite through the continuity of electric current. Observations \citep{Stenbaek-Nielsen2008/JPhD} report glows expanding and merging with beads, sometimes seeding the launch of upward-sideways streamers.  Previous work about the launch of upward streamers in sprites assumed that they emerge from the negative tip of a double-headed streamer \citep{Qin2012/GeoRL,Qin2013/GeoRL}, but this does not seem to be supported by the observations.

Different numerical models indicate that the long-lasting emissions from
glows \citep{Luque2010/GeoRL,Liu2010/GeoRL} and beads \citep{Luque2011/GeoRL}
stem from a persistent electric field and current. \citet{Luque2016/JGRA} suggested
the attachment instability \citep{Douglas-Hamilton1973/ApPhL,Douglas-Hamilton1974/JAP}
as the mechanism responsible for glows and beads. Simulations
show the formation of glow-like structures at altitudes above which the electric field can excite the attachment instability. The situation for beads is still unclear:
they often appear as periodic patterns in observations. \citet{Luque2016/JGRA}
proposed that they emerge from pre-existing horizontal modulations of the atmosphere.

Based on their shape, sprites are classified as carrot-sprites or column-sprites \citep{Wescott1998/JASTP, Stenbaek-Nielsen2008/JPhD}. The main difference is that electrical discharge fronts in column sprites propagate only downwards whereas carrot sprites also shoot upward-sideways streamers in a second stage of evolution.
The inception of these streamers is related to the evolution of the inner electric field in the streamer body
and therefore it depends on the transport of charge and the dominant plasma chemistry.

This paper is organized as follows: in section \ref{sec:model} we describe the 3D  model used to study sprite glows and the emergence of upward streamers. In section \ref{sec:observations} we provide observations of a carrot sprite. In section \ref{sec:results&discussion} we present and discuss our simulation results and compare them with the observation. We finally give some conclusions and new ideas to advance knowledge.

\section{Sprite streamer model}\label{sec:model}
To study the evolution of the upward streamer and the glow itself we use a 3D streamer model. We describe the time evolution of the electron density $n_e$, and ion densities $n_i$, with the drift-diffusion-reaction equations coupled to Poisson's equation:

\begin{linenomath*}
\begin{subequations}
\label{model_eq}
\begin{equation}
\frac{\partial n_{e}}{\partial t}=C_{e}+\nabla\cdot\left(n_e\mu_e \mathbf{E} + D_{e}\nabla n_{e}\right)
+ S_{ph},\label{eq:electrons}
\end{equation}
\begin{equation}
\frac{\partial n_{i}}{\partial t}=C_{i} + S_{ph}~\delta_{O^{+}_{2},i},\label{eq:ions}
\end{equation}
\begin{equation}
-\nabla \cdot \mathbf{E} = \nabla^{2}\phi=-\sum_{s}\frac{q_{s}n_{s}}{\epsilon_{0}},\label{eq:poisson}
\end{equation}
\end{subequations}
\end{linenomath*}
where $i$ indexes the ion species, $s$ runs over ions as well as electrons, $\mu_{e}$ is the electron mobility, $D_{e}$
the electron diffusion coefficient,  $\mathbf{E} $ the electric field,  and $C_{e,i}$ are the net productions of electrons and ions due to plasma-chemical reactions.
We consider ions as motionless over time scales of interest (a few milliseconds).  We use the local field approximation and consequently, transport coefficients are derived from the electron energy distribution function (EEDF) that depends only on the local electric field. $\delta$ is the Kronecker delta and $S_{ph}$ is the photoionization term. Here the Kronecker delta is a symbolic way to express that the photoionization term applies only to $\mathrm{O}_2^+$ in the set of equations \ref{eq:ions}. As for Poisson's equation, $\phi$ is the
electrostatic potential, the sum extends over all the charged species $s$, $q_s$ is the charge of species $s$ and $\epsilon_0$ is the vacuum permittivity.

Regarding the chemistry, we have tested two kinetic schemes. The first one \citep{Luque2016/JGRA} includes electrons, $\mathrm{N}_2$, $\mathrm{O}_2$, $\mathrm{O}_2^+$, $\mathrm{N}_2^+$ and $\mathrm{O}^-$ and
accounts for impact ionization and dissociative attachment. The second one is an extended version with more species and some
other processes such as electron detachment. A detailed description of this second model can be found in the supplementary material of
 \citet{Luque2017/PSST}. Whereas the first model agrees well with observations, the second does not reproduce the long lasting emissions from glows and beads and the exponential decay of the channel reported by observations (a similar problem was discussed by \citet{Luque2016/JGRA}).  Hence, we only present the results of the simpler chemical scheme.

To validate our results against observations, we computed the electron impact excitation of nitrogen molecules to the \NtwoB and \NtwoC electronic states which radiate in the first and second positive systems of $\mathrm{N}_2$ (for details, see the supporting information of \citet{Malagon-Romero2019/GeoRL}).

Our starting point is a single sprite streamer channel right after the streamer head passage. Thus,
our initial condition consists of a vertical neutral column. Detailed streamer simulations show that 1) the radial electron density profile
follows a truncated parabola \citep{Luque2014/NJPh} and that 2) $n_e \sim n_{air}$
when the streamer runs into increasing air density \citep{Luque2010/GeoRL}.
As the streamer propagates downwards, branching or not, it transports positive charge to
the lower region that leads to the decay of the sprite. To model this
we add neutral electron-ion densities in the shape
of a truncated funnel attached to the bottom of the column. Then, the initial electron density reads \citep{Luque2016/JGRA}
\begin{linenomath*}
\begin{equation}
   n_{e,ch}\left(r,z\right) = \begin{cases}
     n_{e0,ch}\left(z\right)\frac{a_0^2}{a\left(z\right)^2}
     \text{max}\left(0,1-\frac{r^2}{a\left(z\right)^2}\right), ~ \forall r
     & ~\SI{50}{km} < z < \SI{80}{km} \\
     0 & \text{elsewhere},
     \label{eq:channel_density}
   \end{cases}
 \end{equation}
\end{linenomath*}
where $n_{e0,ch}\left(z\right) = C n_{air}\left(z\right)$ is the electron density at the axis of the column,  $n_{air}\left(z\right) = n_{air,0} e^{\left(-z/\SI{7.2}{km}\right)}$  is the air density scaled at altitude $z$ from the ground air density $n_{air,0}=\SI{2.5e25}{m^{-3}}$ and $C=\SI{3.33e-11}{}$. The value of $C$ is determined from the equality $n_{e0,ch}\left(\SI{60}{km}\right) = \SI{2e11}{m^{-3}}$. The factor $a_0^2/a(z)^2$ ensures that the total electron density integrated in a horizontal cross section extends smoothly from the funnel to the column. The term $a_0$ is the radius of the column and $a(z)$ is a continuous piece-wise linear function modeling the radius of the channel and the funnel. The funnel is characterized by the radius of the base (\SI{5}{km}), its altitude  ($z_b=\SI{50}{km}$) and the vertex altitude ($z_u = \SI{60}{km}$):
\begin{linenomath*}
\begin{equation}
\label{eq:channel_radius}
 a\left(z\right) = \text{max}\left(a_0,\frac{z_u-z}{z_u-z_b}\times \SI{5}{km}\right)
\end{equation}
\end{linenomath*}

Outside the channel our background electron density follows a Wait-Spies profile \citep{wait1964book}, $n_{e,bg}=\SI{e-2}{cm^{-3}}\exp{\left(-\frac{z-\SI{60}{km}}{\SI{2.86}{km}}\right)}$ \citep{Hu2007/JGRD}.

For the results that we present in this paper, the computational domain is a $20 \times 20 \times 40~\si{km^3}$ box starting at $\SI{40}{km}$ altitude and the finest resolution is $\SI{1}{m}$. To solve Poisson's equation, we ground the bottom boundary and set the top boundary to $\SI{2.9}{MV}$. This is equivalent to a background electric field of $\SI{72.38}{V/m}$ pointing downwards, that yields a reduced electric field $E/n_{air}$ of approximately $\SI{120}{Td}$ at around $\SI{77}{km}$. For the lateral boundaries we set homogeneous Neumann boundary conditions. We also apply Neumann boundary conditions at the
six boundaries to solve electron and ions equations.

Our 3D model is implemented in Afivo-streamer code \citep{Teunissen2017/JPhD,bagheri2019effect}, which is based on the Afivo framework \citep{Teunissen2018/CoPhC}. This framework includes geometric multigrid methods to solve Poisson's equation, octree-based adaptative mesh refinment and OpenMP parallelism. Plasma-chemical reactions and variations in the air density are newly added in Afivo-streamer code for the present simulations.

Afivo-streamer code also includes a 2D cylindrically symmetric version of the model explained above. We have used this 2D version as a probe for suitable conditions to launch upward streamers. These tests show that a sufficiently high vertical electric field suffices to launch upward conical ionization waves. In 3D, this implies that many streamers would emerge from the glow since there is no preferred direction. Under these conditions, our 3D simulations are unapproachable because the extent of the finest grid would be too large. Therefore, we add a small lateral component to the background electric field to break the symmetry.
By doing this we are able to study the propagation of single upward streamer, although more streamers emerge at later stages of our simulations. We have tested several lateral electric fields, however we present the results of a representative simulation, $\mathbf{E_{background}} = \left(18,0,-72.38\right) \si{V/m}$.
The initial (neutral) charge distribution will rapidly evolve to screen the background electric field.

One typical 3D simulation, where the finest grid cell size is \SI{1}{m}, took 3 days in 32 cores. We have checked the convergence of our results by systematically increasing the resolution in our simulations, down to a grid cell size of 0.5 m. Practically the same results were obtained for a grid cell size at and below \SI{1}{m}. Therefore, we just show results calculated for \SI{1}{m} grid spacing in this paper.

\begin{figure*}[h]
\centering
  \includegraphics[width=1.0\textwidth]{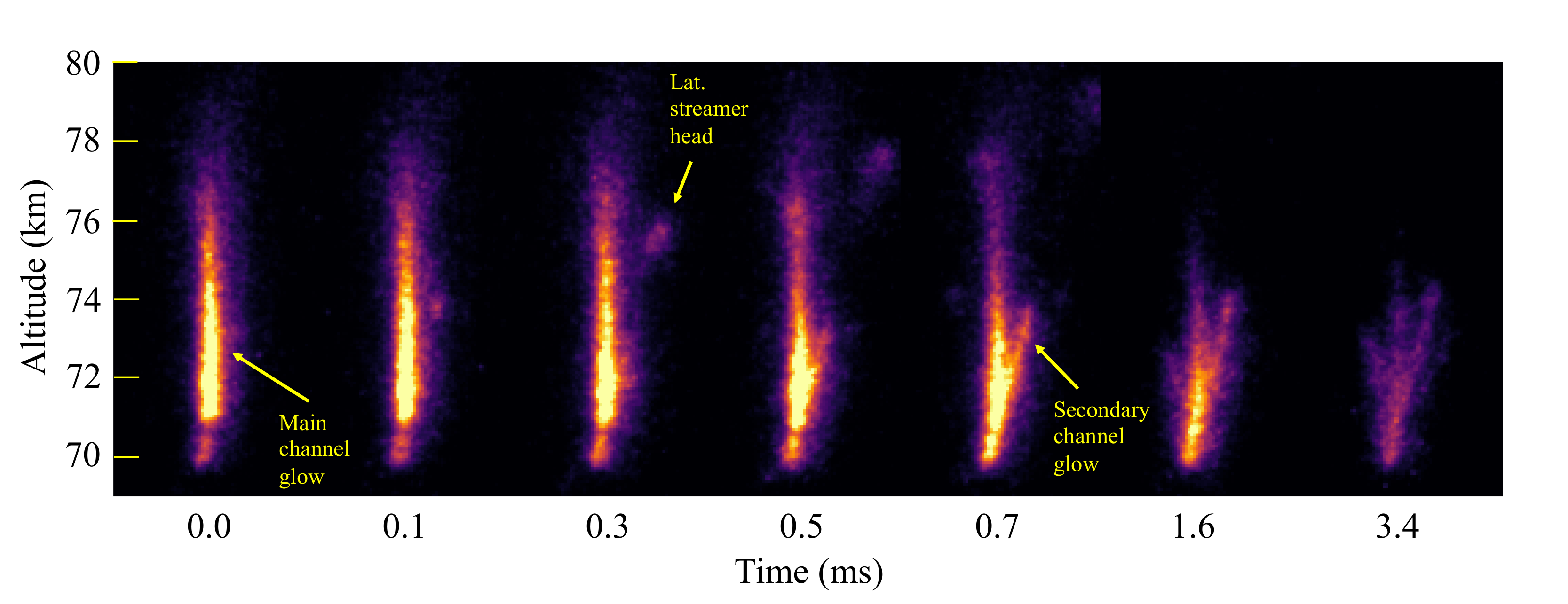}
  \caption{Time series from high-speed video observations of a sprite. The sequence shows the sprite glow and the upward streamer launch, propagation and the optical extinction of the glow.  The time origin was set in the first frame prior to the launch of the upward streamer.  To emphasize the main milestones of the evolution, intervals between the snapshots are not uniform. \label{fig:observations}
    }
\end{figure*}

\section{Observations}\label{sec:observations}
We also present observations of the emergence of an upward streamer from a glowing segment.  After inspecting multiple high-speed sprite observations, we selected one event where this process is particularly clear.  Consistent with earlier results by \citet{McHarg2007/GeoRL}, in all these observations we did not find any event where it could be definitely concluded that an upward streamer did \emph{not} emerge from a glow.

The selected event occurred on 5 July 2011 at 08:54:14 UTC and was associated with a large thunderstorm complex over north-eastern South Dakota, USA. The event was observed with a Phantom high-speed imager recording at 10,000 frames per second onboard an aircraft flying at 14.4 km altitude over south-eastern South Dakota about 180 km north of the storm.

Figure~\ref{fig:observations} shows a time sequence of the recorded event with seven 3.29x1.55 degree subsections from the original 15.23x7.54 degree field of view Phantom images. Once the main channel glow is established, a negative streamer launches and propagates upward, causing a noticeable decrease of the light emissions of the upper part of the glow. Also remarkable is the formation of a secondary channel glow as it is highlighted in the figure. A full video of the event is available in the Supplemental Material and more information about this campaign is provided by \citet{Stenbaek-Nielsen2013/SGeo}.

We do not have the location of the event or the causal lightning stroke, but NLDN reported 5 strikes within about 1 second of the event, all in the region within the Phantom field of view. Assuming that the event is over that region, the range of 180 km mentioned above,  defines the altitude scale shown. Because sprites may occur several tens of km from the causal strike (Stenbaek-Nielsen et al., 2008) there is considerable uncertainty on the altitude. This event was observed at an elevation angle of about 18 degrees and an increase in range by 10 km will increase the altitude by 3.3 km.

\begin{figure*}
\centering
\makebox[0pt]{%
 \includegraphics[scale=0.4]{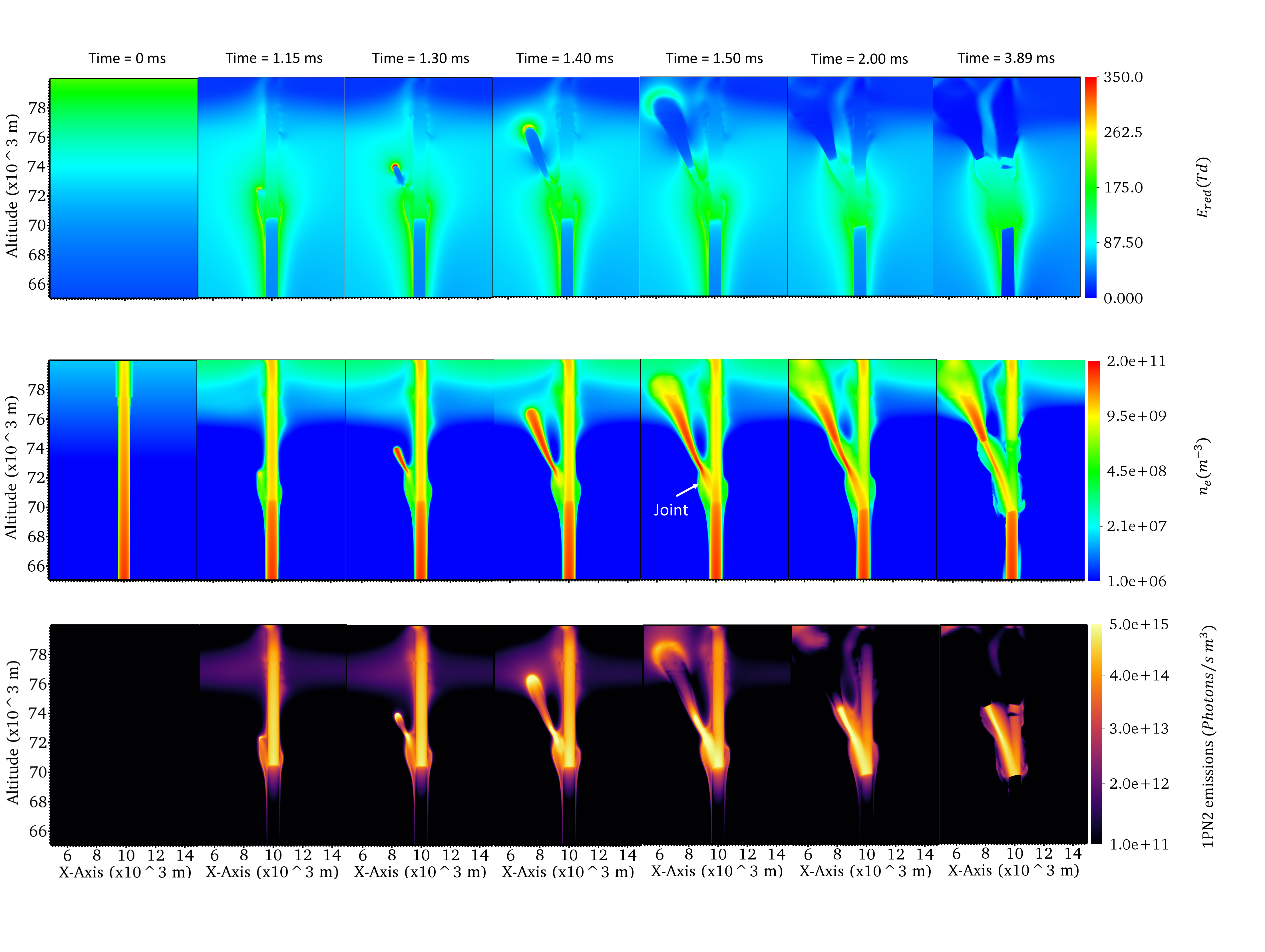}}
  \caption{Cross sections of a 3D simulation in a box domain $20~\si{km} \times 20~\si{km} \times (40-80)~\si{km}$. We show the evolution in terms of the reduced electric field (top row), the electron density (middle row) and the emissions in the First Positive System of molecular nitrogen (1PN2). The channel develops a sharply defined region above \SI{70}{km} characterized by
high reduced electric fields and low conductivity that strongly emits in the 1PN2 for a long time. At $\SI{1.15}{ms}$, a negative streamer emerges close to the lower boundary of the glowing structure, propagates upwards and connects to the electron reservoir in the upper region of the computational domain. The rightmost two columns highlight the effect of the upward streamer on the 1PN2 emissions for altitudes above $\SI{76}{km}$. \label{fig:evolution}
    }
\end{figure*}

\section{Results and Discussion}\label{sec:results&discussion}

Figure~\ref{fig:evolution} shows the evolution of the sprite channel through different milestones: glow formation, upward streamer launch and propagation and, glow deactivation. After \SI{1}{ms} the initial channel appears sharply segmented into high and low conductivity regions below and above $\SI{70}{km}$ respectively. The emissions in the First Positive System of molecular nitrogen (1PN2) (third row in Fig.~\ref{fig:evolution}), can be compared with the recorded event shown in Fig.~\ref{fig:observations}.  In both, simulation and observations, the lower region is almost dark whereas the low conductivity region glows strongly. 1PN2 emissions are determined by the electron density and the reduced electric field. Despite its low conductivity, the upper region emits more intensely because the excitation rate of the 1PN2 is highly non-linear in the reduced electric field. This glowing structure in the upper part of the channel agrees with observations, as the first frame of Fig.~\ref{fig:observations} shows. As \citet{Luque2017/PSST,Malagon-Romero2019/GeoRL} explain, the appearance of this glowing structure is driven by the attachment instability \citep{Douglas-Hamilton1973/ApPhL,Douglas-Hamilton1974/JAP}, which is triggered by relatively high internal electric fields. This is the reason why sprite glows appear in the upper region of the sprite channel \citep{Luque2016/JGRA}.

The boundary of the glow region is characterized by a net negative charge density as previously shown by \citet{Luque2010/GeoRL}.
This is due to current continuity: electrons drift upwards and, around \SI{0.8}{ms} into our simulation, the electron density exceeds the density of positive ions. In the low conductivity region electrons are more effectively converted to negative ions which, being slower than electrons, accumulate there, leading to a surplus of negative charge that locally enhances the electric field. This electric field
has both a vertical and a lateral component and at one point becomes high enough to launch a negative streamer near its lower boundary.  The streamer propagates upwards with a velocity of $\SI{2e7}{m/s}$, typical of streamer observations \citep{Stenbaek-Nielsen2013/SGeo}, and connects to the electron reservoir in the lower ionosphere (see Figure \ref{fig:evolution}). This is very similar to previous \citep{Pasko2002/GeoRL} and our results (fourth frame in Fig. \ref{fig:evolution} ).
Other works \citep{Li2011/GeoRL,Bor2013/JASTP} show upward streamers bending towards the main streamer channel. In our simulation, the upward streamer propagation is mostly straight because we apply a small lateral field  that charges negatively the side of the glow closer to the negative upward streamer, which is slightly repelled.

\begin{figure*}
  \includegraphics[width=1.0\textwidth]{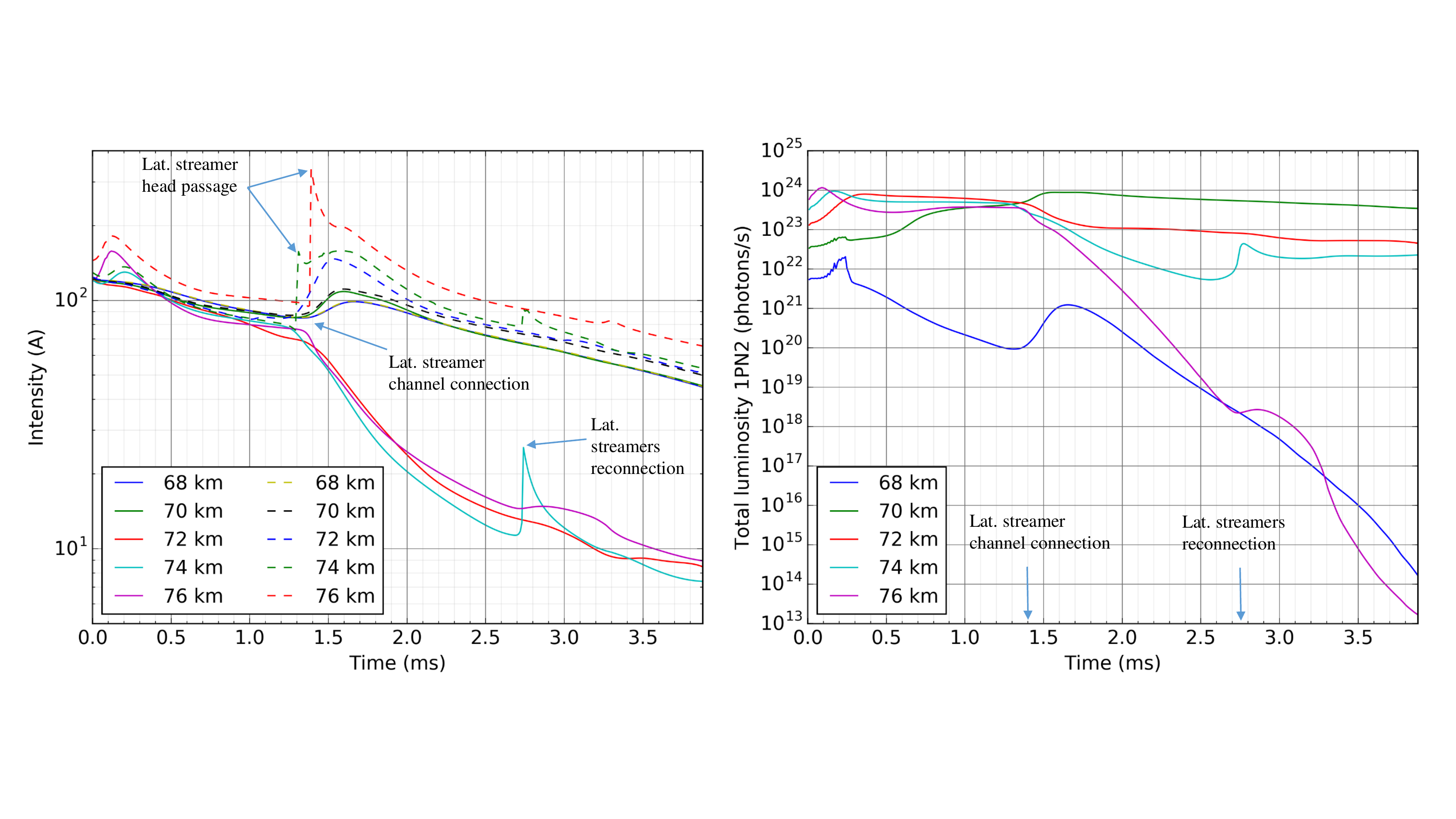}
  \caption{The left plot shows the evolution of the electric current flowing through the main streamer channel (solid lines) and the whole domain (dashed lines). Sudden peaks in the current reveal the passing of the upward streamer head at that altitude.  At around \SI{1.4}{ms} the upward streamer channel and the main streamer channel get effectively connected cutting off the current flowing in the main channel above $\SI{70}{km}$. Subsequent peaks in the current at different altitudes are caused by two negative streamers that emerge from the glow and reconnect to the main channel. The plot to the right shows the emissions from the 1PN2 integrated over a 1km-height box. It clearly shows the effect of the upward streamer, specially at $\SI{76}{km}$. Nontheles, there is also a decrase in the luminosity for $\SI{72}{km}$ and $\SI{74}{km}$.  \label{fig:streamer_effects}
    }
\end{figure*}

The left plot of Fig. \ref{fig:streamer_effects} shows the evolution of the electric current at different altitudes according to the follwoing equation:

\begin{equation}
I=\int_{\Omega=\lbrace r=\left[0,R\right],z=h\rbrace}\mathbf{j}\cdot d\mathbf{s}=2\pi\int_{0}^{R}j_z\left(r,z=h\right)rdr,
\end{equation}

where $j_z$ is the axial component of the current density at $z=h$, r is the radial coordinate (cylindrical coordinates) and R is the radial extent of the
integration domain. Solid lines represent the electric current flowing through the main sprite channel ($R=\SI{400}{m}$) while dashed lines are the current flowing through cross sections spanning the full lateral extent of the domain ($R=\SI{20}{km}$). Initially in the main streamer channel (solid lines) the electric current at different altitudes quickly converges ($\sim \SI{0.5}{ms}$) to the same value despite the initial differences in the background electron density, air density and reduced electric field for each altitude. Afterwards ($\sim \SI{1.4}{ms}$), the upward streamer sets two regions below and above $\SI{70}{km}$. In the whole system (dashed lines), we see sudden current increases at $\SI{74}{km}$ and $\SI{76}{km}$. The peaks for those altitudes at $\SI{1.3}{ms}$ and $\SI{1.4}{ms}$ are due to the passing of the upward streamer head
\citep{Luque2017/PSST}.

As visible in Fig.~\ref{fig:evolution}, the launch point of the upward streamer is slightly separated from the main channel.  The resulting gap or joint initially has a low electron density but around $\SI{1.4}{ms}$, this joint between the upward streamer and the main channel becomes more conductive (see Fig. \ref{fig:evolution}) and, as a result, the currents at $\SI{68}{km}$ and $\SI{70}{km}$ slope upward.
Right after, most of the current starts to flow through the secondary streamer channel, since its higher electron density allows for a higher conductivity, and within the next $\SI{0.3}{ms}$ and for altitudes above $\SI{70}{km}$ the current halves inside the central channel, which has lower electron density. In the system composed of both the main channel and the upward streamer channel, Kirchhoff's current law is a reasonable approximation despite not being strictly applicable to this finite-conductivity, non-steady-state system. The subsequent decay of the currents at different altitudes seems to obey a similar decay constant. This decay is due to the transport of charges inside the streamer channel that increasingly screens the background electric field. Also noticeable is a current peak around 2.7-2.8 ms at $\SI{74}{km}$. This is caused by two negative upward streamers that symmetrically emerge from the main channel, propagate upwards and reconnect to the main channel. We show this event in the supplemental material (movieS2).

1PN2 emissions are useful to track the evolution of the main streamer channel and the impact of the secondary channel on it. In the right panel of Fig. \ref{fig:streamer_effects} we plot the evolution of the 1PN2 emissions integrated over \SI{1}{km}-height and \SI{800}{m}-wide boxes centered at the channel axis. The emissions at $\SI{68}{km}$ rapidly drop due to the screening of the electric field. At other altitudes, specially at $\SI{76}{km}$, the effect of the upward streamer is clear after the rise in conductivity in joint between the upward streamer and the main channel (see Fig. \ref{fig:evolution}). Above $\SI{70}{km}$ the upward streamer induces a reduction of the light emissions whereas below $\SI{70}{km}$ the emissions slightly increase.  When the upward streamer effectively connects (joint becomes conductive enough) to the main streamer channel, the joint glows strongly (at around 1.5 ms  in Fig. \ref{fig:evolution}). In the observations of Fig.~\ref{fig:observations}, this occurs around 0.3-0.5 ms. Later, the emissions in the upper section of the glow decay noticeably. The optical structure of the remnant emissions in the last two frames reproduce what we see in our simulations (Fig.~\ref{fig:evolution}), where most of the emissions come from the lower part of the glow and the region of the upward streamer channel closer to it.

The last stage in the evolution of the upward streamer in our simulation reveals an interesting result ( Fig. \ref{fig:evolution}, last column). The secondary streamer channel develops a sharply defined glowing similar to the glow in the main channel. This portion of the secondary channel has entered the attachment instability regime at around \SI{1.4}{ms}. As a result, the upper boundary of this structure is positively charged and, by a mechanism similar to the one we have described above, it may launch a positive downward streamer.  The resulting optical emissions in that case would feature a vertical channel accompanied at its side by an inverted V. Although infrequently, this formation has indeed been observed and is called ``angel'' sprite \citep{Pasko2012/SSRv,Bor2013/JASTP}.

As mentioned above, we tested two chemical models. We discarded one including detachment because it cannot reproduce long-lasting glows beyond the detachment time scale around $\SI{2}{ms}$. An extensive discussion about the implications of detachment on the glow emissions can be found in \citet{Luque2016/JGRA}. According to other simulations that we do not provide here, detachment delays the upward streamer but does not prevent it as long as its onset occurs before $\SI{2}{ms}$ (in our simulations) .

The morphology of the sprite is related to the electric field at mesospheric altitudes. We tested electric fields with smaller vertical components, and we found that for lower
electric fields the glow is shorter. This means that the upward streamer, if it were to appear, would emerge later. If the glow is too small to
accumulate enough charge at the lower boundary, it is unlikely that it launches upward streamers. Column-sprites, which are associated to less intense electric fields \citep{Qin2013/JGRA,Qin2013/GeoRL}, would hardly launch upward streamers.

\section{Conclusions}

Our simulations show for the first time the evolution of a sprite streamer channel where a glow region forms in its upper part and launches several upward streamers. We have proposed a mechanism for the emergence of carrot sprites: the sprite glow results from an attachment instability that seeds the lower boundary of the glow with negative charge that locally enhances  the electric field and finally launches an upward negative streamer.
The evolution of the upward streamer and the glow itself agrees well with observations for the time scales we have simulated.
Essentially, the upward streamer cuts off the current flowing from the lower part of the sprite channel to the glow region, which then darkens. This agrees with many observations where the glow gets dimmer once the upward streamers are emitted. Of course, the details depends on the strength and number of upward streamers. In their last stage our simulations reveals a region in the upward streamer that mimics the glowing structure in the main channel. We have seen that positive charge accumulates in its upper boundary. This new region might be the origin of positive downwards streamers which would give rise to the ``angel'' sprite structure \citep{Pasko2012/SSRv,Bor2013/JASTP}. Observations rarely report ``angel'' sprites and in our simulations, we have not succeeded in reproducing them. Therefore, its production must be constrained to a small subset in our parameter space.

Finally, there is still a puzzling question concerning the electron detachment in sprite simulations. It is a process that we should consider to study the streamer propagation for sufficiently long times. However, results point out that detachment should somehow be suppressed or treated in a different way. At atmospheric pressure the presence of a sufficiently high water concentration effectively suppresses the electron detachment \citep{Malagon-Romero2019/GeoRL}. Noctilucent clouds reveal the presence of water at sprite altitudes. Whether this suffices to affect detachment noticeably is still unknown. Nevertheless, water is a candidate to be considered in future work.

\acknowledgments
Information on how to access the code used to run the simulations as well as the output data analyzed in this study is available in the supporting information.
This work was supported by the European Research Council (ERC) under
the European Union H2020 programme/ERC grant agreement 681257. A. Malagón-Romero and A. Luque acknowledge financial support from the State Agency for Research of the Spanish MCIU through the "Center of Excellence Severo Ochoa" award for the Instituto de Astrofísica de Andalucía (SEV-2017-0709). J. Teunissen was supported by postdoctoral fellowship 12Q6117N from Research Foundation -- Flanders (FWO). The aircraft observations were sponsored by the Japanese National Broadcasting System, NHK. H.C. Stenbaek-Nielsen and M. G. McHarg were supported by grants 1104441 and 1201683 from the US National Science Foundation to the University of Alaska Fairbanks and the US Air Force Academy, respectively. The simulation code used for this work is available at \url{https://gitlab.com/MD-CWI-NL/afivo-streamer}. The version used for the simulations corresponds to the commit a3e2241fba15f177b26e79106cef7354cbe09d68. Configuration files as well as the output data are available at \url{https://doi.org/10.5281/zenodo.3478177}.

\newcommand{\jgr}{J. Geophys. Res.}
\newcommand{\grl}{Geophys. Res. Lett.}
\newcommand{\physrep}{Phys. Rep.}
\newcommand{\mdash}{---}

\end{document}